\documentclass[final,authoryear,5p]{elsarticle}

\usepackage{epsfig}
\usepackage{graphicx}
\usepackage{caption}
\usepackage{subcaption}

\usepackage{amssymb}

\usepackage[ps2pdf,%
a4paper=true,%
breaklinks=true,%
colorlinks=true,%
pdfauthor={Senthamizh Pavai et al.},%
pdftitle={Tilt angle measurements from historical sunspot observations}%
]{hyperref}

\journal{Advances in Space Research}

\begin{document}

\begin{frontmatter}



\title{Sunspot group tilt angle measurements from historical 
observations
}


\author{V. Senthamizh Pavai, R. Arlt\corref{cor}
}
\address{Leibniz Institute for Astrophysics Potsdam (AIP), An der Sternwarte 16, 14482 Potsdam, Germany}
\cortext[cor]{Corresponding author}
\ead{svalliappan@aip.de}


\author{A. Diercke
}
\address{AIP; Institut f\"ur Physik und Astronomie, Universit\"at Potsdam,
Karl-Liebknecht-Str. 24/25, 14476 Potsdam, Germany}

\author{C. Denker
}
\address{Leibniz Institute for Astrophysics Potsdam (AIP), An der Sternwarte 16, 14482 Potsdam, Germany}

\author{J. M. Vaquero
}
\address{Dept. de F\'{\i}sica, Centro Universitario de M\'erida, Universidad de Extremadura, Avda Santa Teresa de Jornet 38, 06800 M\'erida, Spain}

\begin{abstract}
Sunspot positions from various historical sets of solar drawings 
are analysed with respect to the tilt angles of bipolar sunspot 
groups. Data by Scheiner, Hevelius, Staudacher, Zucconi, Schwabe, 
and Sp\"orer deliver a series of average tilt angles spanning a 
period of 270 years, additional to previously found values for
$20^\mathrm{th}$-century data obtained by other authors. We find that the
average tilt angles before the Maunder minimum were not 
significantly different from the modern values. However, the
average tilt angles of a period 50~years after the Maunder minimum, namely
for cycles~0 and~1, were much lower and near zero. The normal 
tilt angles before the Maunder minimum suggest that it was not
abnormally low tilt angles which drove the solar cycle into a
grand minimum.

\end{abstract}


\end{frontmatter}

\parindent=0.5 cm

\section{Introduction}

The long-term study of solar cycle properties using historical observations
has provided mainly the sunspot number \citep{clette_ea2014}. However, we
can also access other properties more directly related to the solar dynamo
through historical sunspot drawings. The tilt angle of sunspot 
groups is among those parameters and can be included in certain types of 
dynamo models. Bipolar sunspot groups exhibit an axis through the two
main magnetic polarities. The tilt angle is the angle at which this axis 
is orientated with respect to the solar equator. It is an important property 
in flux-transport dynamos \citep[Babcock-Leighton dynamos, see e.g.]
[Sect. 4.8]{charbonneau2010} in which it provides the 
source term for the poloidal magnetic field which in turn correlates with the 
strength of the next cycle. The tilt angles are widely believed to be the result
of buoyantly unstable magnetic flux tubes at the bottom of the convection
zone, rising under the influence of rotation, internal twist, and magnetic
tension. 

According to results of the thin flux tube approximation, 
tilt angles are either due to writhing of rising flux 
loops by the Coriolis force or the pitch angle of the 
subsurface field wound up by the differential rotation 
\citep{dsilva_choudhuri1993}. The combination of magnetic 
buoyancy and the Coriolis force generates the correct latitudinal 
distribution of tilt angles, according to numerical simulations in the thin flux tube framework
\citep[e.g.][]{dsilva_choudhuri1993,caligari_ea1995,fan_fisher1996,weber_ea2013}. 
These computations can also reproduce the correlation between the magnetic field 
strength and the tilt angle which is seen in some observational studies 
\citep{tian_ea2003,dasi_ea2010}. The average tilt angle and the amplitude 
of the corresponding cycle appears to be anti-correlated, while the 
product of the average tilt angle with the cycle amplitude is well
correlated with the strength of the following cycle \citep{dasi_ea2010,dasi_ea2013}.

\begin{table*}
\caption{Mean and median tilt angles for various data sets. Full widths at half-maximum (FWHM) were derived from Gaussian fits to the tilt angle distributions.}
\begin{tabular}{lcrrrr}
\hline\\[-3mm]
Data source& Years      &\multicolumn{1}{c}{Mean tilt}&\multicolumn{1}{c}{Median tilt}& FWHM  & Number of groups \\[1mm]
\hline\\[-3mm]
Scheiner   & 1618--1627 & $3.92^\circ \pm 0.91^\circ$ & $ 3.63^\circ \pm 0.91^\circ$  & $33.4^\circ $ & 537    \\
Hevelius   & 1642--1644 & $4.79^\circ \pm 1.43^\circ$ & $ 5.35^\circ \pm 1.43^\circ$  & $27.1^\circ $ & 130    \\
Staudacher & 1749--1796 & $2.03^\circ \pm 0.98^\circ$ & $ 1.79^\circ \pm 0.98^\circ$  & $50.3^\circ $ & 828    \\
Zucconi    & 1754--1760 & $0.62^\circ \pm 2.28^\circ$ & $-1.48^\circ \pm 2.28^\circ$  & $35.8^\circ $ & 131    \\
Schwabe    & 1825--1867 & $4.45^\circ \pm 0.20^\circ$ & $ 4.69^\circ \pm 0.20^\circ$  & $40.6^\circ $ & 15548  \\
Sp\"orer   & 1861--1894 & $4.86^\circ \pm 0.45^\circ$ & $ 4.04^\circ \pm 0.45^\circ$  & $31.3^\circ $ & 2834   \\[1mm]
\hline
\label{avgtilt}
\end{tabular}
\end{table*}

In thin flux tube models, the tilt angles are even useful in constraining 
the strength of initial magnetic flux. The strength of the toroidal magnetic 
field at the bottom of the convection zone has to be in the range of 
$40$--$50$~kG in order to obey the observed Joy's law 
\citep{weber_fan_miesch2011}.

Observational studies by \citet{kosovichev_stenflo2008} show that the tilt 
angles of sunspot groups change gradually over their lifetime
except in the beginning of emergence. While the tilt angles are random
in the earliest phase of emergence, they adjust towards Joy's law
during the rest of the emergence phase, i.e.\ as long as the magnetic
flux is growing. It is not straight-forward to draw a direct link between
the tilt angles and the emergence of flux tubes in simulations.
The average tilt angles are also fairly independent of the cycle 
phase within fixed latitudinal zones \citep{li_ulrich2012}.

The study of tilt angles derived for several centuries helps us  
understand their origin and their relation to the solar cycle. The 
true tilt angles of sunspot groups are available only from magnetic
data of the solar surface for the second half of the $20^\mathrm{th}$ century, 
while pseudo-tilt angles are measured without the polarity information 
and have a $180^\circ$ ambiguity. Pseudo-tilt angles
can be computed whenever individual spot positions in sunspot groups
are available from drawings or images. They have recently been calculated 
for the period of 1825--1867 using the sunspots observations by Schwabe 
\citep{schwabe_2015}. 

In this paper, we present the tilt angle measurements from further historical 
sunspot observations, namely the observations by Christoph Scheiner (1618, 1621--1622, 
1625--1627), Johannes Hevelius (1642--1644), Johann Caspar Staudacher 
(1749--1796), Ludovico Zucconi (1754--1760), and Gustav Sp\"orer (1861--1894).

The details of different solar observations and the methods used in data 
extraction from those sunspot drawings are described in Sect.~\ref{data_obs}. 
The comparison of mean tilt angles and cycle-mean tilt angles from various 
data are discussed in Sect.~\ref{tilt}.

\section{Data set}\label{data_obs}
Christoph Scheiner started his sunspot observations from Ingolstadt, Germany, in 
the early 17th century. His first known sunspot drawing was made on 21~October
1611. Most of his data, however, were recorded from Rome.
He published observations only for a few days during each of the years 
of 1611, 1612, 1618, 1621, 1622, and 1624. In the period 1625--1627, the 
observations are fairly continuous (drawings covering 342~days in 1625, 
163~days in 1626, and 55~days in January--June 1627). His drawings show 
the sunspot groups traversing the solar disk in a single full-disk drawing \citep{scheiner_1630}.
The positions and areas of the sunspots were measured using 13~circular cursor 
shapes with areas between one and 364~pixels. The data before 1618 were not included 
in the tilt angle distribution, because they are extremely coarse and show
highly exaggerated sizes of sunspot groups.

Johannes Hevelius recorded his observations of the Sun from Gda\'nsk, Poland, 
during the period of 1642--1644 (15 days in October--December 1642, 
110 days in May--December 1643, and 98 days in January--October 1644). 
These sunspot drawings were published by Hevelius in an appendix
of his book {\it Selenographia\/} \citep{hevelius_1647}.
His drawing style is very similar to Scheiner's style, and the positions 
and area information were obtained in the same way. It is important to 
note that these sunspot drawings were made just before the Maunder
Minimum, a period of reduced solar activity from 1645 to 1715 approximately 
\citep{spoerer1889,usoskin_ea2015}.

During the second half of the 18th century Johann Caspar Staudacher (or Staudach) recorded 
his observations of the Sun from Nuremberg, Germany, and made drawings in 
1749--1796. Detailed information about the drawings and the data extraction 
methods can be found in \citet{Arlt_2008} and \citet{Arlt_2009}. Various
methods of estimating the orientation of the drawings had to be employed
to measure the sunspot positions. From the data base derived, we use only 
the spots with quality flags~1 and~2. This basically excludes drawings 
for which the orientation was estimated using a typical tilt angle for bipolar regions.
Quality-3 observations use the tilt as an input in many cases and are therefore
not used.

For a rather short time from April 1754 to May 1758 and a short spell in 
June 1760, Ludovico Zucconi observed the Sun from Venice, Italy, contemporaneously 
with Staudacher. The positions and areas of individual sunspots were extracted 
by \citet{hsunspots_2011} using the HSUNSPOTS tool. The orientation of
these drawings were clearly marked by the observer, and we consider them
fairly precise.

\begin{figure} 
\centering
\includegraphics*[width=8.7cm]{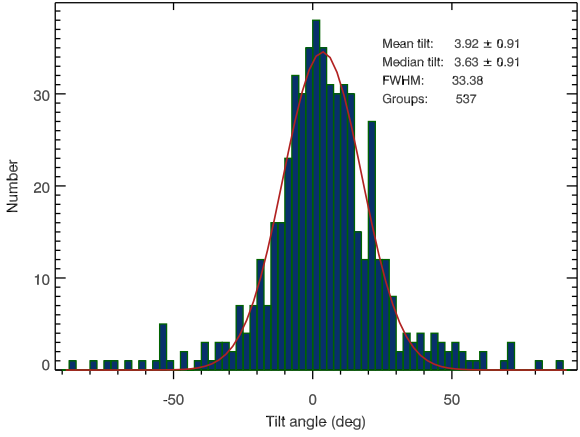}
\caption{Histogram of the tilt angles of sunspot groups from the drawings
by Scheiner. Only groups with area weighted centers within $\pm60^\circ$ CMD 
are used. The FWHM was derived from a Gaussian fit (solid line).}
\label{tilt_schein}
\end{figure}

\begin{figure} 
\centering
\includegraphics*[width=8.7cm]{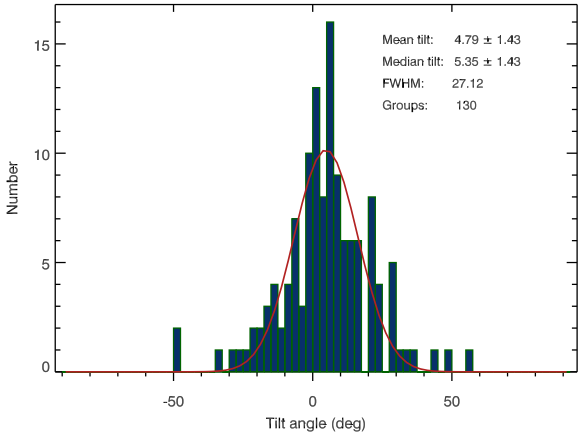}
\caption{{Tilt angle histogram for Hevelius. 
Only groups with area weighted centers within $\pm60^\circ$ CMD  
are used.}}
\label{tilt_hev}
\end{figure}

\begin{figure} 
\centering
\includegraphics*[width=8.7cm]{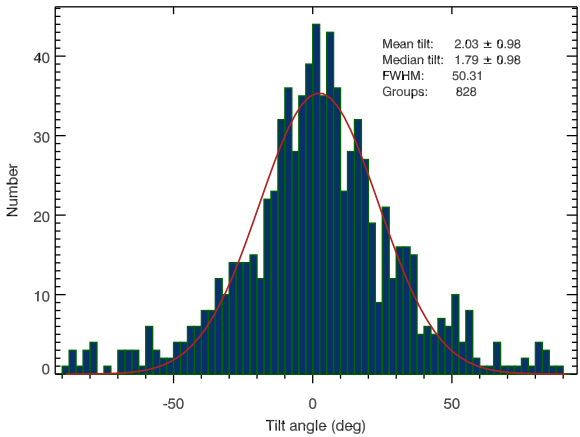}
\caption{{Tilt angle histogram for Staudacher. 
Only groups with area weighted centers within $\pm60^\circ$ CMD, 
polarity separations $\Delta\beta>3^\circ$, and with quality flags 1 and 2 
are used.}}
\label{tilt_stau}
\end{figure}

\begin{figure} 
\centering
\includegraphics*[width=8.7cm]{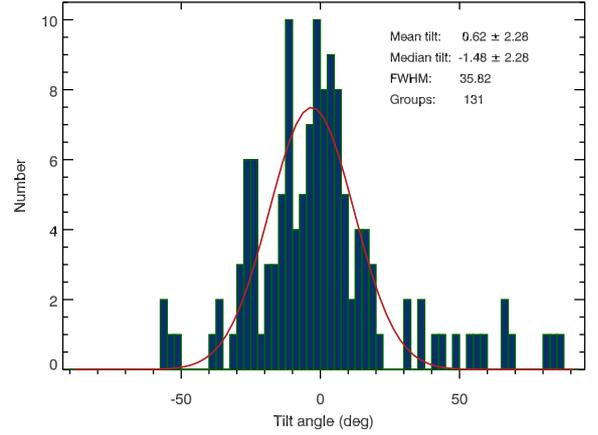}
\caption{{Tilt angle histogram for Zucconi. 
Only groups with area weighted centers within $\pm60^\circ$ CMD  
 are used.}}
\label{tilt_zuc}
\end{figure}

\begin{figure} 
\centering
\includegraphics*[width=8.7cm]{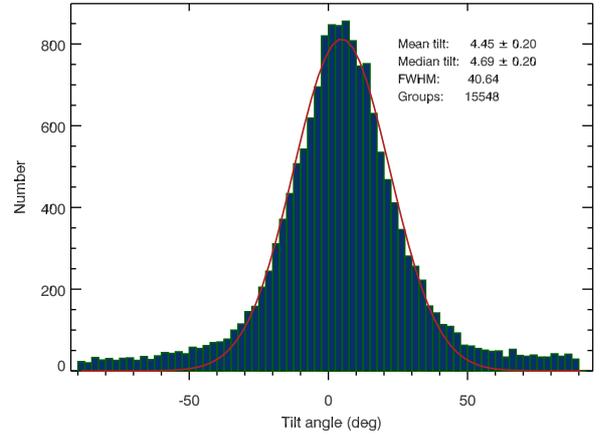}
\caption{{Tilt angle histogram for Schwabe. 
Only groups with area weighted centers within $\pm60^\circ$ CMD and 
polarity separations $\Delta\beta>3^\circ$ are used.}}
\label{tilt_sch}
\end{figure}

\begin{figure} 
\centering
\includegraphics*[width=8.7cm]{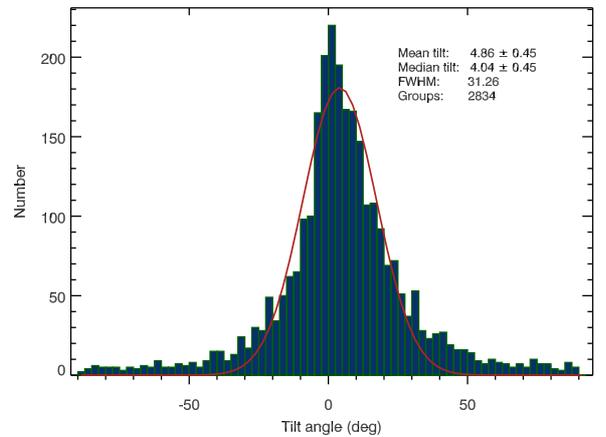}
\caption{{Tilt angle histogram for Sp\"orer and groups with polarity 
separations $\Delta\beta>3^\circ$. Only spots with areas $\geq 1$~MSH 
were considered while calculating the tilt angle.}}
\label{tilt_spor}
\end{figure}

The drawings of sunspot observations made by Samuel Heinrich 
Schwabe from Dessau, Germany, in the period 1825--1867, and the 
extraction of data from them were explained in detail by \citet{arlt2011} 
and \citet{arlt_ea2013}. A description of the method that was employed to compute all the 
tilt angles in the present paper is given by \citet{schwabe_2015}, where
it was applied to the Schwabe data.

Friedrich Wilhelm Gustav Sp\"orer observed sunspots from Anklam and Potsdam, 
Germany, during 1861--1894. He drew the sunspot groups while they crossed 
the central meridian, so the evolution of sunspot groups is not available. 
The details of the drawings and the technique used in the extraction of 
positions and areas of sunspots were given by \citet{Dierckeetal_2015}.

\begin{figure}
\centering
\includegraphics[width=8.7cm]{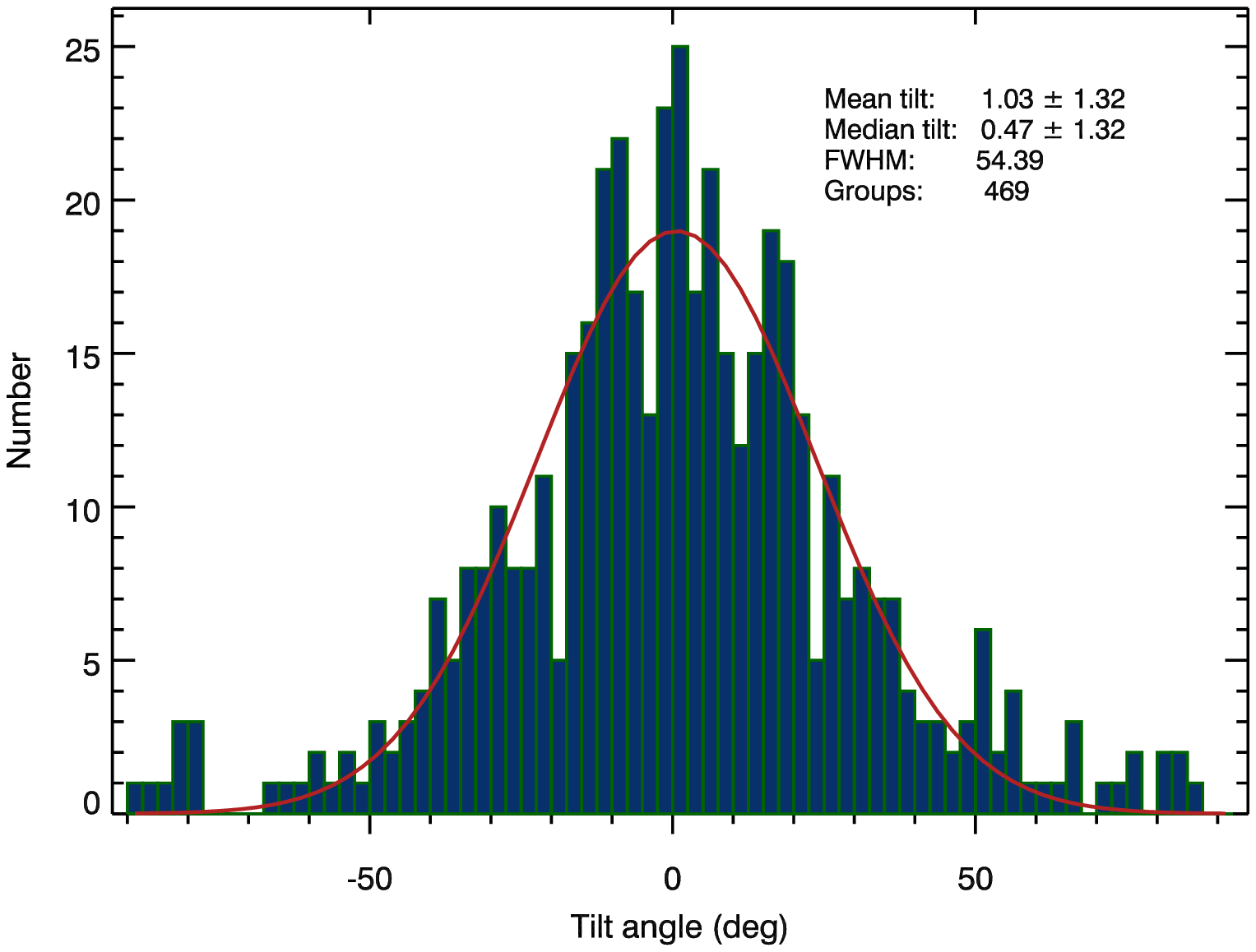}\\
\includegraphics[width=8.7cm]{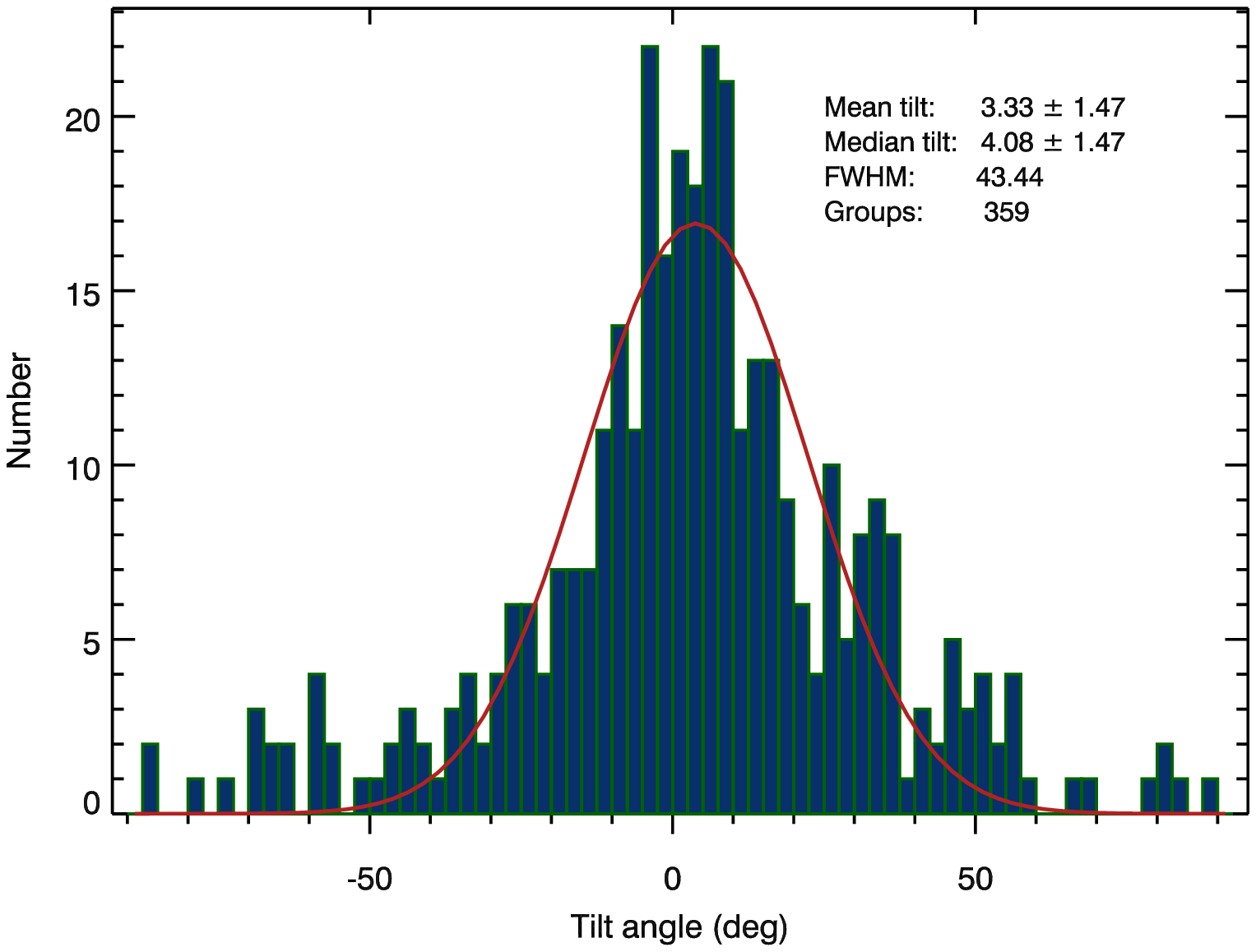}
\caption{{Tilt angle histograms for cycles~0 and~1 (top) and cycles~2, 3 
     and~4 (bottom) from Staudacher. Only groups with area weighted centers 
     within $\pm60^\circ$ CMD, polarity separations $\Delta\beta>3^\circ$ and with 
     quality flags~1 and~2 are used.}}
\label{tilt_stau_part}
\end{figure}

\section{Tilt angle distributions}\label{tilt}
Based on the positions of the individual spots, a tangential plane is adopted
touching the solar surface in the group center in order to minimize
curvature effects in determining the tilt angle. The groups are 
divided into two polarities such that the variance of the spot positions 
in the individual polarities are lowest. 
The method is explained in detail in \cite{schwabe_2015}. The sign 
of the tilt angle is positive when the leading polarity is closer to the 
equator than the following polarity. There is no sign of the polarities 
available from the sunspot drawings. Therefore, groups not obeying Hale's 
polarity law cannot be detected.

The definition of what forms a sunspot group varies among the various
observers. We have therefore inspected all the data sets used here to
re-group sunspots where it appeared necessary. We used the spot distributions
as well as the evolution of the groups to discriminate them. The following 
numbers refer to groups through their full life-time as a single group, not 
to individual appearances of groups in each drawing. While the re-grouping 
for Schwabe is described in \citet{schwabe_2015}, a number of groups were
split or combined in the records of the other observers as well. This
procedure led to 13~new groups in Scheiner's drawings, whereas in seven
cases, groups were combined to one. In the drawings by Hevelius, we obtained
five new groups and combined groups in three cases. Zucconi's drawings
contained two groups which needed to be split into two new groups each.
Finally, 104~new groups were formed in Sp\"orer's data, while in 90~cases,
two groups were combined to a single group. Since Staudacher did not provide
any group designations, the association of spots to groups was made
from scratch by inspection of the drawings.

From those sunspot groups, the bipolar ones need to be extracted for the computation 
of tilt angles. The sunspot drawings of Scheiner, Hevelius, and Zucconi were 
manually inspected to select the bipolar groups, and only those groups are 
included in the present study. Figures~\ref{tilt_schein}, \ref{tilt_hev}, 
and~\ref{tilt_zuc} show the distributions of tilt angles from the drawings 
of Scheiner, Hevelius, and Zucconi, respectively. In the distributions, 
only the bipolar groups within $\pm60^\circ$ central meridian distance 
(CMD) are included, because the positional accuracy drops 
significantly beyond these limits. In all the graphs, we fit the discrete 
distributions with Gaussian functions to obtain estimates for the full 
widths at half-maximum (FWHM).

In case of Staudacher, Schwabe, and Sp\"orer (Figs.~\ref{tilt_stau}, 
\ref{tilt_sch}, and~\ref{tilt_spor}, respectively), the volumes of 
sunspot drawings were too large to manually pick the bipolar groups. 
For Staudacher and Schwabe, we complemented the condition of 
limiting the groups within $\pm60^\circ$ CMD by a
minimum polarity separation ($\Delta\beta$) of $3^\circ$ \citep{baranyi2015} 
in order to statistically remove unipolar groups. Figure~\ref{tilt_spor} 
shows the tilt angle distribution of sunspot groups from Sp\"orer's 
observations. Since the groups in Sp\"orer's drawings are already 
at central meridian, only the condition $\Delta\beta> 3^\circ$ was 
applied to remove potentially unipolar groups. The averages and medians of 
all the above tilt angle distributions are given in Table~\ref{avgtilt}.

The FWHM for the tilt angle distributions are 
in the range of $30^\circ$--$40^\circ$, with one exceptional value of 
$50^\circ$ for Staudacher. \citet{wang_ea2015} obtained an FWHM of 
$30.8^\circ$ for the distribution of tilt angles from the Debrecen data using 
umbral data alone, as we do here as well.

\begin{table*}
\caption{Average and median tilt angles for individual solar cycles or 
small groups of cycles as well as the FWHM 
of the individual distributions. We use the cycle number notation of 
\citet{zolotova_ponyavin2015}. Parentheses indicate that the cycle was
not fully covered by observations.}
\begin{tabular}{llcccr}
\hline\\[-3mm]
Data source& Cycle & Mean tilt & Median tilt  & FWHM  & Groups \\[1mm]
\hline\\[-3mm]
Scheiner   & $-12$ (first half)&($3.92^\circ \pm 0.91^\circ$)&($3.63^\circ \pm 0.91^\circ$)  &($33.4^\circ$) &  537  \\
Hevelius   & $-10$ (maximum)   &($4.79^\circ \pm 1.43^\circ$)&($5.35^\circ \pm 1.43^\circ$)  &($27.1^\circ$) &  130  \\
Zucconi    & 0+1 (min. between 0 and 1) &($0.62^\circ \pm 2.28^\circ$)&($-1.48^\circ\pm 2.28^\circ$)  &($35.8^\circ$) &  131  \\
Staudacher & 0+1               & $1.03^\circ \pm 1.32^\circ$ & $0.47^\circ \pm 1.32^\circ$   & $54.4^\circ$  &  469  \\
           & 2+3+4             & $3.33^\circ \pm 1.47^\circ$ & $4.08^\circ \pm 1.47^\circ$   & $43.4^\circ$  &  359  \\
Schwabe    &  7 (second half)  &($3.24^\circ \pm 0.55^\circ$)&($3.12^\circ \pm 0.55^\circ$)  &($41.0^\circ$) & 2243  \\
           &  8                & $4.36^\circ \pm 0.47^\circ$ & $5.05^\circ \pm 0.47^\circ$   & $42.9^\circ$  & 3419  \\
           &  9                & $4.71^\circ \pm 0.34^\circ$ & $4.79^\circ \pm 0.34^\circ$   & $40.3^\circ$  & 4942  \\
           & 10                & $4.74^\circ \pm 0.36^\circ$ & $4.94^\circ \pm 0.36^\circ$   & $39.7^\circ$  & 4898  \\
Sp\"orer   & 10 (end)          &($2.91^\circ \pm 1.15^\circ$)&($1.77^\circ \pm 1.15^\circ$)  &($19.5^\circ$) &  465  \\
           & 11                & $4.59^\circ \pm 0.76^\circ$ & $4.59^\circ \pm 0.76^\circ$   & $32.0^\circ$  & 1067  \\
           & 12                & $5.73^\circ \pm 0.80^\circ$ & $4.63^\circ \pm 0.80^\circ$   & $34.3^\circ$  &  833  \\
           & 13 (beginning)    &($5.85^\circ \pm 1.11^\circ$)&($5.64^\circ \pm 1.11^\circ$)  &($32.6^\circ$) &  469  \\[1mm]
\hline
\label{cycleavgtilt}
\end{tabular}
\end{table*}

The sizes of the sunspots in Staudacher's drawings are highly exaggerated, 
so the area values were not used in the tilt angle calculation. Sp\"orer 
magnified the sizes of sunspots in his drawings to some extent, and we
needed to scale down the area values by a factor of 13.3 as inferred 
by \citet{Dierckeetal_2015}. Sp\"orer recorded both pores and 
umbrae but other drawings only contain umbrae or umbrae and penumbrae. To 
make it consistent with other data, only umbrae with areas $\geq 1$~MSH were 
considered while calculating the tilt angle.

\subsection{Comparison of cycle-averaged tilt angles}
The cycle-averaged tilt angle is a quantity related to the 
polar field generated by active regions in the course of a cycle 
and may be an indication of the activity of the future cycle. 
These averages are available since solar cycle~15 from modern data 
\citep[e.g.][]{dasi_ea2010,mcclintock_norton2013,wang_ea2015}. 
Since historical data provide tilt angles over a much longer period, we look 
at cycle-averaged tilt angle values from Staudacher, Schwabe, and Sp\"orer 
data which cover the cycles~0--4 and~7--13. The data by Scheiner and Hevelius 
are not covering entire solar cycles. Their average may only be a rough
indication of the cycle-average tilt angle and need to be treated with
caution.

The Staudacher data covers solar cycles~0--4 without the beginning of cycle~0. 
Table~\ref{avgtilt} shows that the mean tilt in the Staudacher 
and Zucconi data are lower than the ones from other data sources. While 
Staudacher's drawings are not very precise and could contain a strong 
random component bringing the average tilt angle close to zero, Zucconi's 
data are precise enough and confirm the very low average tilt. The data by
Zucconi cover the cycle minimum between solar cycles~0 and~1 \citep[February 1755
according to][]{hathaway_2010} showing spots
at latitudes from $0^\circ$ to $30^\circ$, apparently including spots from 
both the ceasing cycle~0 and the growing cycle~1. 

The butterfly diagram from the Staudacher data for the solar cycles~0 and~1 
also shows a peculiar behaviour with an excess of groups at the equator \citep{Arlt_2009}. 
We therefore divide the Staudacher data into two parts, such that one part 
contains the cycles~0 and~1 and the other part contains the remaining data 
(cycles~2--4), instead of dividing the data into individual cycles. The data 
do not contain the beginning of the cycle~0 but together with cycle~1, we
consider the result sufficiently representative for a two-cycle average. The 
mean tilt angles were found to be $1.03^\circ\pm1.32^\circ$ for cycles~0 and~1 
and $3.33^\circ\pm 1.47^\circ$ for cycles~2--4. Figure~\ref{tilt_stau_part} shows
the tilt angle distributions separately for cycles~0 and~1 and the remaining part of
the data. Using only the data with 
quality flag of 1 (disk orientation obtained from rotational matching of spots
on adjacent days), the distinction between the two types of cycles is even clearer:
the mean tilt for cycles~0 and~1 is $0.24^\circ \pm 1.61^\circ$, and it is 
$5.68^\circ \pm 1.83^\circ$ for the cycles~2--4. The mean tilt angle for the cycles 
0 and 1 is lower and on par with the mean tilt of Zucconi's data. The 
combined mean tilt for cycles~2--4 from the Staudacher data are higher
and compatible with values of modern data. 
While the Staudacher data alone may be too inaccurate, the agreement with
Zucconi's data is an indication for the peculiarity of the cycles~0 and~1.

The Schwabe data comprise cycles~7--10, while the Sp\"orer data cover 
cycles~10--13. The data for the initial two~years of cycle~7 are missing
in the Schwabe data, however. Data from Sp\"orer are also not available for the first 
half of cycle~10 and the second half of cycle~13. The spots being located at predominantly very low latitudes
at the end of cycle~10 cause an underestimate of the mean tilt angle, according 
to Joy's law. Table~\ref{cycleavgtilt} lists the mean and median tilt values 
for a number of cycles in the period $\sim$1620--1890. The FWHM were 
again derived from fits with Gaussian distributions.
The average tilt for cycles~0 and~1 are the lowest of all solar cycles for
which we have analysed drawings so far.

\section{Conclusions}
The various historical sunspot observations by Scheiner, Hevelius, Staudacher, 
Zucconi, Schwabe, and Sp\"orer offer white-light sunspot drawings during the 
$17^\mathrm{th}$, $18^\mathrm{th}$, and $19^\mathrm{th}$ centuries. The tilt 
angles for the supposedly bipolar sunspot groups from those different sunspot observations 
were calculated. The values from the period before the Maunder minimum 
(years in the 1620s and 1640s) are comparable to precise $20^\mathrm{th}$-century 
results. The importance of the tilt angles for the transport of magnetic
flux on the surface and the polar field was first noticed by \citet{leighton1964}.
Various effects of averaged tilt angles have been studied 
more recently \citep[cf.\ e.g.][]{baumann_ea2004, cameron_ea2010}.
The fact that the pre-Maunder minimum average tilt angles are relatively 
large suggests that it was not particularly low values that initiated
a period of very low activity. If flux-transport dynamos are indeed
operating in the Sun, more subtle effects such as, e.g., the occurrence
of equator-crossing groups \citep{cameron_ea2013} or other group properties
need to be studied to find precursors for the very deep minimum following cycle~$-10$.

The cycle-averaged tilt angle values were also calculated for the solar 
cycles 0--4 and 7--13 from Staudacher, Schwabe and Sp\"orer data. The mean tilt 
value for the cycles~0 and~1 seems to be the lowest of all cycle-averaged tilt 
angles, independently shown from sunspot drawings by Staudacher and precise
images made by Zucconi. The accuracy of the Staudacher images may be 
questioned on the one hand, especially their orientation, but the 
agreement with Zucconi's data, on the other hand, is striking. The 
Sun also appeared to have a slightly stronger differential rotation 
in that period as compared to today \citep{Arlt_frohlich2012}, but 
the result was not significant. Since this period is about three to 
four cycles after the Maunder minimum, it is not clear whether the
features still represent the recovery from a grand activity minimum.
While that early Staudacher period needs careful future inspection, 
the cycles before and after the Maunder minimum are clearly very 
valuable in constraining which kind of dynamo is operating in the Sun.


\section*{Acknowledgements}
VSP acknowledges support by grant AR 355/10-1, CD acknowledges support 
by grant DE 787/3-1, both of the German Science Foundation (DFG). JMV
acknowledges support by grants GR15137 and AYA2014-57556-P.






\begin{thebibliography}{}


\bibitem[Arlt(2008)]{Arlt_2008}
Arlt, R., Digitization of sunspot drawings by Staudacher in 1749--1796,
SoPh, 247, 399--410, 2008.

\bibitem[Arlt(2009)]{Arlt_2009}
Arlt, R., The Butterfly Diagram in the Eighteenth Century, SoPh, 255, 143--153, 2009.

\bibitem[Arlt (2011)]{arlt2011}
Arlt, R., The sunspot observations by Samuel Heinrich Schwabe, Astron. Nachr.,
332, 805--814, 2011.

\bibitem[Arlt \& Fr\"ohlich(2012)]{Arlt_frohlich2012}
Arlt, R., Fr\"ohlich, H.-E., The solar differential rotation in the 18th 
century, A\&A, 543, A7, 6pp., 2012.

\bibitem[Arlt et al.(2013)]{arlt_ea2013}
Arlt, R., Leussu, R., Giese, N., Mursula, K., Usoskin, I.G., Sunspot 
positions and sizes for 1825--1867 from the observations by Samuel 
Heinrich Schwabe, MNRAS, 433, 3165--3172, 2013.

\bibitem[Baranyi(2015)]{baranyi2015}
Baranyi, T., Comparison of Debrecen and Mount Wilson/Kodaikanal sunspot 
group tilt angles and the Joy's law, MNRAS, 447, 1857--1865, 2015.

\bibitem[Baumann et al.(2004)]{baumann_ea2004}
Baumann, I., Schmitt, D., Sch\"ussler, M., Solanki, S.K., Evolution of the 
large-scale magnetic field on the solar surface: A parameter study,
A\&A, 426, 1075--1091, 2004.

\bibitem[Caligari et al.(1995)]{caligari_ea1995}
Caligari, P., Moreno-Insertis, F., Sch\"ussler, M., Emerging flux tubes
in the solar convection zone. I. Asymmetry, tilt, and emergence latitude,
ApJ, 441, 886--902, 1995.

\bibitem[Cameron et al.(2010)]{cameron_ea2010}
Cameron, R.H., Jiang, J., Schmitt, D., Sch\"ussler, M., Surface flux
transport modeling for solar cycles 15--21: effects of cycle-dependent
tilt angles of sunspot groups, ApJ, 719, 264--270, 2010.

\bibitem[Cameron et al.(2013)]{cameron_ea2013}
Cameron, R.H., Dasi-Espuig, M., Jiang, J., I\c{s}\i{}k, E., Schmitt, D., Sch\"ussler, M.,
Limits to solar cycle predictability: Cross-equatorial flux plumes, A\&A,
557, A141, 6pp., 2013.

\bibitem[Charbonneau(2010)]{charbonneau2010}
Charbonneau, P., Dynamo models of the solar cycle, Living Rev. Solar Phys., 7, 3--91, 2010.

\bibitem[Clette et al.(2014)]{clette_ea2014}
Clette, F., Svalgaard, L., Vaquero, J.M., Cliver, E.W., Revisiting the sunspot number.
A 400-year perspective on the solar cycle, Space Sci. Rev. 186, 35--103, 2014.

\bibitem[Cristo et al.(2011)]{hsunspots_2011}
Cristo, A., Vaquero, J.M., S\'anchez-Bajo, F., HSUNSPOTS: a tool for the 
analysis of historical sunspot drawings, JASTP, 73, 187--190, 2011.

\bibitem[Dasi-Espuig et al.(2010)]{dasi_ea2010}
Dasi-Espuig, M., Solanki, S.K., Krivova, N.A., Cameron, R., Pe{\~n}uela, T., 
Sunspot group tilt angles and the strength of the solar cycle, A\&A, 518, A7, 10pp., 2010.

\bibitem[Dasi-Espuig et al.(2013)]{dasi_ea2013}
Dasi-Espuig, M., Solanki, S.K., Krivova, N.A., Cameron, R., Pe{\~n}uela, T.,
Sunspot group tilt angles and the strength of the solar cycle (corrigendum), A\&A, 556, C3, 2013.

\bibitem[Diercke et al.(2015)]{Dierckeetal_2015}
Diercke, A., Arlt, R., Denker, C., Digitization of sunspot drawings by 
Sp\"orer made in 1861--1894. Astr. Nachr., 336, 53--62, 2015.

\bibitem[D'Silva \& Choudhuri(1993)]{dsilva_choudhuri1993}
D'Silva, S., Choudhuri, A.R., A theoretical model for tilts of bipolar 
magnetic regions, A\&A, 272, 621--633, 1993.

\bibitem[Fan \& Fisher(1996)]{fan_fisher1996}
Fan, Y., Fisher, G.H., Radiative heating and the buoyant rise of
magnetic flux tubes in the solar interior, SoPh, 166, 17--41, 1996.

\bibitem[Hathaway(2010)]{hathaway_2010}
Hathaway, D.H., The solar cycle, Living Rev. Solar Phys., 7, 1--65, 2010.

\bibitem[Hevelius(1647)]{hevelius_1647}
Hevelius, J., Selenographia sive, lun\ae{} descriptio, H\"unefeld, Gda\'nsk, 563~pp., 1647

\bibitem[Kosovichev \& Stenflo(2008)]{kosovichev_stenflo2008}
Kosovichev, A.G., Stenflo, J.O., Tilt of emerging bipolar magnetic regions 
on the Sun. ApJL, 688, L115--L118, 2008.

\bibitem[Leighton(1964)]{leighton1964}
Leighton, R.B., Transport of magnetic fields on the Sun, ApJ, 140, 1547--1562, 1964.

\bibitem[Li \& Ulrich(2012)]{li_ulrich2012}
Li, J., Ulrich, R.K., Long-term measurements of sunspot magnetic tilt angles, 
ApJ, 758, 115, 12pp., 2012.

\bibitem[McClintock \& Norton(2013)]{mcclintock_norton2013}
McClintock, B.H., Norton A.A., Recovering Joy's law as a function of solar
cycle, hemisphere, and longitude, SoPh, 287, 215--227, 2013.

\bibitem[Scheiner(1630)]{scheiner_1630}
Scheiner, C., Rosa ursina sive sol, Andrea Fei, Bracciano, 822~pp., 1630.

\bibitem[Senthamizh Pavai et al.(2015)]{schwabe_2015}
Senthamizh Pavai, V., Arlt, R., Dasi-Espuig, M., Krivova, N.A., Solanki, S.K., 
Sunspot areas and tilt angles for solar cycles 7--10, A\&A, 584, A73, 13~pp., 2015.

\bibitem[Sp\"orer(1889)]{spoerer1889}
Sp\"orer, G., Ueber die Periodicit\"at der Sonnenflecken seit dem Jahre 1618. In: Nova Acta der Ksl. Leop.-Carol. Deutschen Akademie der Naturforscher, Vol.~8, No.~2, 283--324, 1889.

\bibitem[Tian et al.(2003)]{tian_ea2003}
Tian, L., Liu, Y., Wang, H., Latitude and magnetic flux dependence of the 
tilt angle of bipolar regions, SoPh, 215, 281--293, 2003.

\bibitem[Usoskin et al.(2015)]{usoskin_ea2015}
Usoskin, I.G., Arlt, R., Asvestari, E., Hawkins, E., K\"apyl\"a, M. et al., 
The Maunder minimum (1645--1715) was indeed a grand minimum: A reassessment 
of multiple datasets, A\&A, 581, A95, 19~pp., 2015.

\bibitem[Wang et al.(2015)]{wang_ea2015}
Wang, Y.-M., Colaninno, R.C., Baranyi, T., Li, J., Active-region tilt angles: 
magnetic versus white-light determinations of Joy's law, ApJ, 798, 50, 14~pp., 2015.

\bibitem[Weber et al.(2011)]{weber_fan_miesch2011}
Weber, M.A., Fan, Y., Miesch, M.S., The rise of active region flux tubes in
the turbulent solar convection zone, ApJ, 741, 11, 14~pp., 2011.

\bibitem[Weber et al.(2013)]{weber_ea2013}
Weber, M.A., Fan, Y., Miesch, M.S., Comparing simulations of rising flux tubes
through the solar convection zone with observations of solar active regions:
constraining the dynamo field strength, SoPh, 287, 239--263, 2013.

\bibitem[Zolotova \& Ponyavin(2015)]{zolotova_ponyavin2015}
Zolotova, N.V., Ponyavin, D.I., The Maunder minimum is not as grand as it
seemed to be, ApJ, 800, 42, 14~pp., 2015.

%

\end{thebibliography}
\end{document}